\documentclass[copyright,creativecommons]{eptcs}
\renewcommand{\event}{S. Barry Cooper \& Vincent Danos (eds.):\\
Fifth Workshop on Developments in Computational Models\\
---Computational Models From Nature}
\providecommand{\volume}{??}
\providecommand{\anno}{2009}
\providecommand{\firstpage}{1}
\usepackage{breakurl}

\usepackage{amsmath,amsfonts,algorithm}
\usepackage{graphicx}
\usepackage{amssymb}
\usepackage{latexsym}
\usepackage{color}
\usepackage{bbm}
\usepackage[all]{xy}\CompileMatrices

\newcommand{\odd}{\textup{Odd}} 

\def\enco{A}
\def\qed{$\Box$}

\def\ie{\textit{i.e.}}

\def\al{\alpha}

\newtheorem{prop}{Proposition}\def\PRO{\begin{prop}}\def\ORP{\end{prop}}
\newtheorem{coro}{Corollary}\def\COR{\begin{coro}}\def\ROC{\end{coro}}
\newtheorem{theo}{Theorem}\def\TH{\begin{theo}}\def\HT{\end{theo}}
\def\TH{\begin{theo}}\def\HT{\end{theo}}
\newtheorem{defi}[prop]{Definition}\def\DE{\begin{defi}}\def\ED{\end{defi}}
\newtheorem{definition}[prop]{Definition}
\newtheorem{lemme}[prop]{Lemma}\def\LE{\begin{lemme}}\def\EL{\end{lemme}}
\newcommand{\AR}[2][c]{$$\begin{array}[#1]{lllllllllllllll}#2\end{array}$$}

\def\bra#1{\langle#1{|}}
\def\ctR{\mathop{\wedge}\hskip-.4ex} 

\def\cz#1#2{Z_{#1}^{#2}}
\def\cx#1#2{X_{#1}^{#2}}

\def\emi#1{e^{{-i}#1}}
\def\cx#1#2{X_{#1}^{#2}}

\def\M#1#2{{M}_{#2}^{#1}}

\def\al{\alpha}

\def\ket#1{| #1 \rangle}

\def\bra#1{\langle #1 |}

\newcommand{\den}[1]{[\hspace{-0.06cm}[ #1]\hspace{-0.06cm}]}

\definecolor{Ocolor}{rgb}{.9,.1,.1}
\definecolor{Bcolor}{rgb}{.1,.1,.9}

\title{Information Flow in Secret Sharing Protocols}

\author{E. Kashefi \institute{School of Informatics, University of Edinburgh, UK} \email{ekashefi@gmail.com}
\and  D. Markham  \institute{LTCI CNRS, Telecom ParisTech, France} \email{markham@enst.fr}
\and  M.Mhalla \institute {CNRS, Universit\'e de Grenoble, France} \email{mehdi.mhalla@imag.fr}
\and  S.Perdrix \institute{CNRS, Universit\'e Paris Diderot, France} \email{simon.perdrix@gmail.com}
}
\def\titlerunning{Information Flow in Secret Sharing Protocols}
\def\authorrunning{E. Kashefi \& D. Markham \& M.Mhalla \& S.Perdrix}
\begin{document}

\maketitle

\begin{abstract}
The entangled graph states~\cite{HEB04} have emerged as an elegant and
powerful quantum resource, indeed almost all multiparty protocols can be written in terms of graph states including measurement based quantum computation (MBQC), error correction and secret sharing amongst others. In addition they are at the forefront in terms of implementations. As such they represent an excellent opportunity to move towards integrated protocols involving many of these elements. In this paper we look at expressing and extending graph state secret sharing \cite{MS08} and MBQC in a common framework and graphical language related to flow \cite{g-flow,Flow06}. We do so with two main contributions.

First we express in entirely graphical terms which set of players can access which information in graph state secret sharing protocols. These succinct graphical descriptions of access allow us to take known results from graph theory to make statements on the generalisation of the previous schemes to present new secret sharing protocols.

Second, we give a set of necessary conditions as to when a graph with flow, i.e. capable of performing a class of unitary operations, can be extended to include vertices which can be ignored, \emph{pointless measurements}, and hence considered as unauthorised players in terms of secret sharing, or error qubits in terms of fault tolerance. This offers a way to extend existing MBQC patterns to secret sharing protocols. Our characterisation of pointless measurements is believed also to be a useful tool for further integrated measurement based schemes, for example in constructing fault tolerant MBQC schemes.

\end{abstract}

\section{Introduction}

The one-way model is a universal model of quantum computation
introduced by Briegel and Russendorf \cite{RB01} that consists of performing a
sequence of one-qubit measurements on an initial entangled quantum
state. A detailed description of the model is captured in the measurement calculus \cite{DKP}. A pattern $\mathcal P=(V,I,O,A)$, is defined over a set of qubits $V$, where $I/O$ is the subset representing input/output qubits, and $A$ is the measurement calculus commands, listed below. The semantics of a pattern $\den{\mathcal P}:\mathfrak h_I \to \mathfrak h_O$ is defined as follows.
\vspace*{1em}\pagebreak

\begin{eqnarray*}
\den{(V,I,O,A)}&:=&\rho\mapsto \sum_{s\in \{0,1\}^{O^c}} \den{A}_s \rho \den{A}_s^\dagger\\
\den{A'A}_s&:=&\den {A'}_s\circ \den{A}_s\\
\den{E_{i,j}}_s&:=&\rho\mapsto \ctR Z_{i,j}\rho \ctR Z_{i,j}\\
\den{M_i^\alpha}_s&:=&\rho\mapsto \begin{cases}\bra{+_\alpha}_i\rho\ket{+_\alpha}_i&\text{if $s_i=0$}\\\bra{-_\alpha}_i\rho\ket{-_\alpha}_i&\text{if $s_i=1$}\end{cases}\\
\den{X_i^{s_j}}_s&:=&\rho\mapsto \begin{cases}\rho&\text{if $s_j=0$}\\X_i\rho X_i&\text{if $s_j=1$}\end{cases}\\
\den{Z_i^{s_j}}_s&:=&\rho\mapsto \begin{cases}\rho&\text{if $s_j=0$}\\Z_i\rho Z_i&\text{if $s_j=1$}\end{cases}\\
\end{eqnarray*}

The main focus of this paper is the structural characterisation of secret sharing protocols within the framework of measurement-based quantum computing. In doing so we will use the the notation of \emph {flow} that is a very useful tool to chracterize the feasibility of a measurement-based computation using a graph state \cite{Flow06,g-flow} that can be computed in polynomial time \cite{MP08}.
\DE
$(g,\prec)$ is a gflow of $(G,I,O)$, where $g:V(G)\setminus O \to
\wp(V(G)\setminus I)$ and $\prec$ is a strict partial order over $V(G)$, if and only if\\
\indent 1. if $j\in g(i)$ then $i\prec j$\\
\indent 2. if  $j\in Odd(g(i))$ then $j=i$ or $i\prec j$\\
\indent 3. $i \in Odd(g(i))$\\
Where $Odd (K)=  \{u\, ,\,  |N(u)\cap K|=1 \mod 2\}$ is the odd neighbourhood of $K$, i.e. the set of vertices which have an odd number of neighbours in $K$.
\ED

A detailed introduction on secret sharing protocols can be found in \cite{MS08}. Here we present the notations and definitions that are required in this paper.  Informally speaking an $(n,k)$ secret sharing protocol is a multi-partite protocol with a special party called \emph{dealer}, who holds a secret $S$, which is either a bit  or a qubit, and must  send this secret to $n$ players such that any $k$ or more
players can reconstruct the secret, and all sets of
fewer than $k$ players as well as eavesdroppers are
denied any access whatsoever to the secret.

Following \cite{MS08}, this can then be broken down into three separate settings. First, sharing of a classical secret, where there is already a secure channel between the dealer and the players, denoted {\sf CC}. Second, sharing of a classical secret, where there is possible eavesdropping between the dealer and the players and we use quantum channels to over come this, denoted {\sf CQ}. Third sharing of a quantum secret over quantum channels between the dealer and players, denoted {\sf QQ}. Although the {\sf CC} graph state scheme gives no advantage over existing classical schemes, they are very useful to consider as they offer insight into the {\sf CQ} and {\sf QQ} cases where there is a quantum advantage (indeed {\sf QQ} is by definition a quantum problem).

One can formalise a {\sf CC} classical secret sharing protocol over a graph state $G$ as follows. Let $A\subseteq V(G)$, a classical secret sharing protocol $(G,A)$ is:
\begin{itemize}
\item A dealer holds a secret $s\in \{0,1\}$ and prepares the state $$\ket {\phi(s)} = Z_A^s\ket G$$ where $\ket G=\prod_{(u,v)\in E(G)} \ctR Z_{u,v} \bigotimes_{u\in V(G)} \left(\ket 0_u+\ket 1_u\right)/\sqrt 2$
\item The dealer sends each qubit of $\ket {\phi(s)}$ to a distinct player.
\end{itemize}

\begin{definition}
Given a protocol $(G,A)$ and a set $S\subseteq V(G)$ of players, the reduced density matrix associated with these players is:

$$ \rho_S(s):=tr_{V\setminus S}(\ket {\phi(s)}\bra{\phi(s)})$$
\end{definition}


\begin{definition}[Privacy]
A secret is \emph{private} from a set of players if they cannot obtain any information about the secret. For {\sf CC} secret sharing the secret this corresponds to a set $S\subseteq V(G)$  such that $$\rho_S(0)= \rho_S(1).$$
\end{definition}

\begin{definition}[Accessibility] A secret \emph{accessible} to a set $S$ if they can reconstruct the secret perfectly. For {\sf CC} this corresponds to a set $S\subseteq V(G)$ such that $$tr(\rho_S(0)\rho_S(1))=0.$$
\end{definition}

\begin{definition} [Protocol] A secret sharing protocol $(G,A)$ is $k$-\emph{private}, if all sets of fewer than $k$ players cannot obtain any information about the secret. A secret sharing protocol $(G,A)$ is $k$-\emph{accessible}, if any $k$ or more players can reconstruct the secret.

\noindent A protocol $(G,A)$ with $n=|V(G)|$ is a \emph{$(k,n)$- secret sharing protocol} if the protocol is both $k$-private and $k$-accessible.
\end{definition}

The extensions to {\sf CQ} and full {\sf QQ} quantum secret sharing protocols of \cite{MS08} are obtained by using the above graph $G$ and attaching an ancilla qubit to the encoding set $\enco$ which is kept by the dealer. In the dealer versus players partition this can be viewed as a maximally entangled bipartite state, which is used as a quantum channel to establish a secret random key between the dealer and authorised players for {\sf CQ} (a kind of extension of entanglement based key distribution \cite{Ekert91}), or to teleport the secret from the dealer to authorised players for {\sf QQ}. The fact that the players are separate means they are restricted in how they can carry out their half of the protocol to access the secret. The access of information in theses cases is not just governed by the graph $G$, but also the {\it conjugate graph}, defined below  (see \cite{MS08} for more details).

\DE
The conjugate graph $G'$ with respect to graph $G$ and encoding qubits in $\enco$ is defined as the graph given by taking the complementation of graph $G$ over set $\enco$.
\ED



\section{Characterisation}

We present a complete and simple graphical condition as to which sets of players can access the secret. In what follows we define the \emph{encoding qubits}, $\enco$, to be the set of qubits in $G$ where the dealer will act upon them to encode the initial secret.

\TH \label{t-charac1}
For the {\sf CC} classical secret sharing protocols $(G,A)$ of \cite{MS08}, the secret can be accessed by a set $S$ if there exists $D\subseteq S$ such that
\begin{align}
 D \cup Odd(D)\subseteq S \nonumber\\
|D \cap \enco | = 1 \mod 2 \nonumber,
\end{align}
\HT
{{\bf Proof.} The proof consists in introducing an observable $P_S$ acting on the qubits of $S$ such that the quantum state sent by the dealer is commuting with $P_S$ if the secret is $0$ and anticommuting otherwise. Thus the players in $S$ can reveal the secret by measuring their qubits according to $P_S$.

Let $P_S:=\prod_{u\in D}X_u\left(\prod_{v\in N_G(u)}Z_v\right)$. Since $D\cup Odd_G(D) \subseteq S$, $P_S$ acts on $S$. Moreover, $P_S\ket{G}=\ket G$ since $X_u\left(\prod_{v\in N_G(u)}Z_v\right)\ket G=\ket G$ for any $u\in V(G)$. Finally, $P_SZ_\enco \ket{G}= (-1)^{|D\cap A|}Z_\enco P_S\ket G= -Z_\enco \ket G$ since $X$ and $Z$ anticommute and $|D\cap \enco |=1[2]$. Thus, $P_SZ^s_\enco \ket{G}=(-1)^sZ^s_\enco\ket{G}$.
}
\qed

\vspace{0.2cm}

We define $Acc$ as the minimal accessing sets, that is $S\in Acc$ if $S$ can access the information and $\forall v\in S, S\setminus \{v\}$ cannot access the information. Any set that contains an element of $Acc$ can  access the information. Similarly we define $Blk$ as the minimal blocking sets   $S\in Blk$ if $V\setminus S$ cannot access the information and $\forall v\in S, V\setminus( S\setminus\{v\})$ can access the information. Any set that contains an element of $Blk$ can block the information.

\begin{lemme}
$Acc$ and $Blk$ are transversal sets.
\end{lemme}
{\bf Proof.}
 By definition, each contains the minimal elements that has a  non-empty intersection with each member of the other. Indeed if $S$ can access the information then no set in $V\setminus S$ can block it and vice versa.
\qed

We finish with another characterisation result for secret sharing.


\TH \label{t-charac2}
For the {\sf CC} classical secret sharing protocols of \cite{MS08} on graph G, the secret \emph{cannot} be accessed by a set $S$ if there exists $K\in V(G)\setminus S$ such that  $$Odd(K)\cap S = \enco \cap S $$
where $\enco$ denotes the encoding qubits.
\HT
{\bf Proof.} If the condition is satisfied, then the players in $V\setminus S$ can act on the state such that the reduced state of the players in $S$ does not depend on the secret. Indeed, let $P := \prod_{u\in K}X_u\left(\prod_{v\in N_G(u)\setminus S}Z_v\right)$. By definition, $P$ is acting on $V\setminus S$. Moreover $\ket G$ is a fixpoint of $PZ_{A\cap S}$, so $Z_A\ket G = Z_{A\setminus S}P\ket G$. Thus the reduced density matrix of the players in $S$ does not depend on the secret $s\in \{0,1\}$: $\rho_S(1) = tr_{V\setminus S}(Z_A\ket {G}\bra{G}Z_A) =  tr_{V\setminus S}(Z_{A\setminus S}P\ket {G}\bra{G}P^\dagger Z_{A\setminus S}) = tr_{V\setminus B}(\ket {G}\bra{G}) = \rho_B(0)$.
\qed\\

The conditions for access in the {\sf CQ} and {\sf QQ} cases are the same, but should be satisfied by both $G$ and the conjugate graph $G'$ defined in the last section \cite{MS08}. That is, sets that can access for both $G$ and $G'$ (for the same encoding set $\enco$), where access is determined by Theorems 1 and 2, can access the secret when extended to {\sf CQ} and {\sf QQ}. Further, sets that cannot access information in the {\sf CC} case for $G$ cannot access the secret for the extension to {\sf CQ} and {\sf QQ}.

Clearly from the proofs, one can verify that Theorem \ref{t-charac1} implies Theorem \ref{t-charac2}. In fact it can also be shown that implication holds the otherway around also \cite{Simon09}. Thus, either the classical information can be accessed locally, by stabilisers, in accordance with Theorem 1 or it cannot be accessed at all, and the condition of Theorem 2 is satisfied. Since these Theorems also govern the manipulation of quanutm information via the conjugate graph, the implications are also important for error correction and likely can be extended to other means of pushing around quantum information. This is potentially very powerful tool to try and prove many statements about information access and manipulation in graph sates, and will be presented in other work \cite{Simon09}. Here we will concentrate on the relevance to secret sharing.

\section{New Protocols and other implications}

Despite the simple condition of theorems \ref{t-charac1} and \ref{t-charac2}, it is still an open question which graphs will satisfy them. However these theorems allow us to take known results from graph theory to make statements on the generalisation of the secret sharing schemes presented in \cite {MS08} as we describe here.

We consider different cases for the graph $G=(V,E)$, where the access set can be obtained based on the results of the previous section. In theses examples all the qubits are encoding qubits, $\enco=V$.

The first case we consider is the complete bipartite graphs. We split the partners in two parties depending on the side of the bipartition they are in, and we prove that to access the information the unanimity from one side of the partition and at least a partner from the second side are required.

\TH
If $G$ is a complete bipartite graph $G=K_{n,n}$ with $n\ge 2$. Then the accessing set for {\sf CC} secret sharing is defined as follows.
\AR{
Acc(G)=\{\{u,v_1\ldots,v_n \} \in V,(u,v_i)  \in E \}
}
\HT
{\bf Proof.}
An accessing  set  is of the form $D \cup Odd(D)$ with $|D|$ odd. Thus $D$ contains an odd number of vertices in one side of the partition which implies that the other side is in $Odd(D)$. Minimal sets of that form contains only one vertex in one side of the partition.
\qed

\begin{figure}[!h]
\begin{center}
\includegraphics[width=5cm]{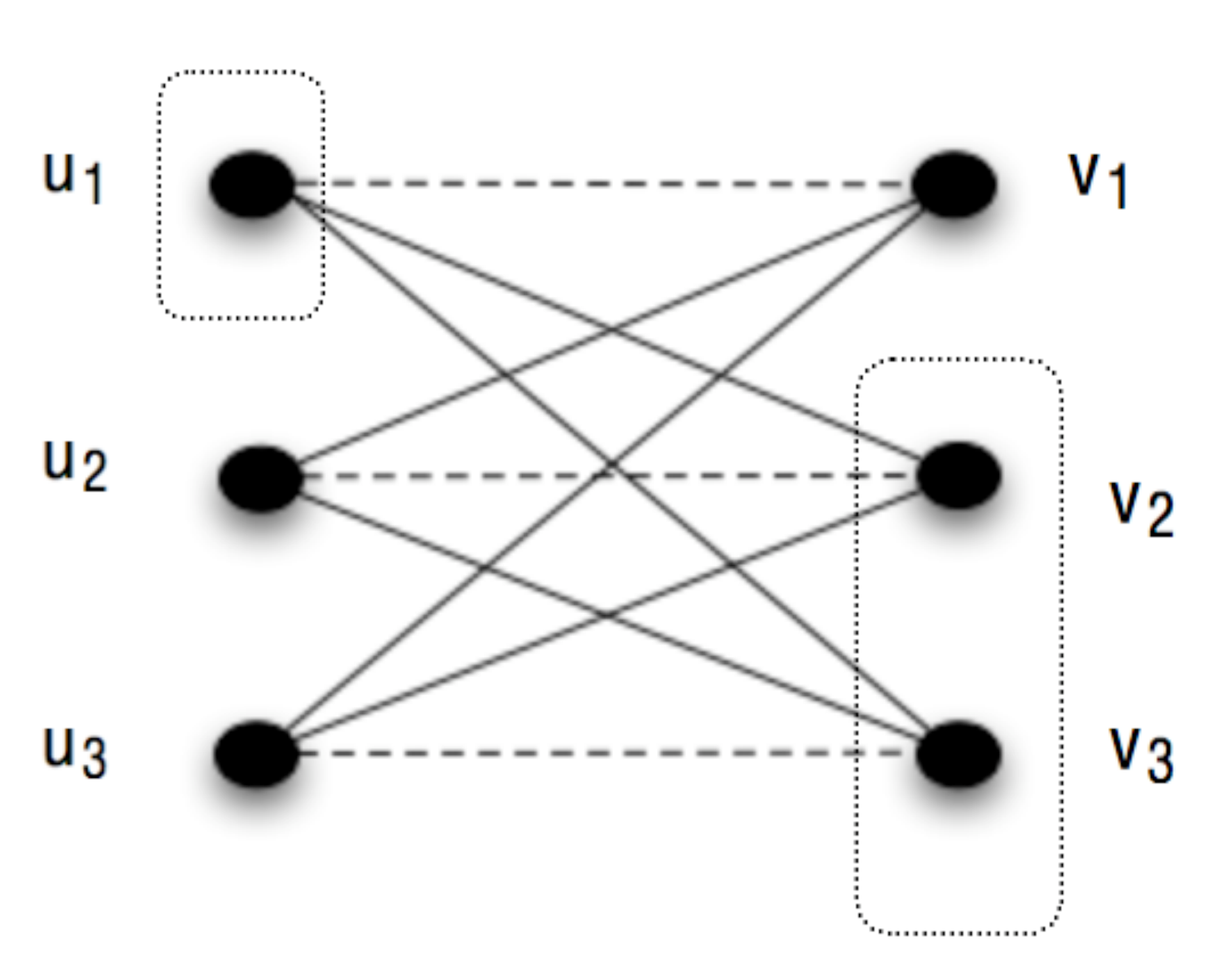}
\caption{Secret sharing with a bipartite graph minus a matching}
\label{bipm}
\end{center}
\end{figure}

The second case is when $G$ is obtained from the complete bipartite graph by removing a perfect matching $M$ (Figure \ref{bipm}). A perfect matching is a set of edges that intersects each vertex once. We split the participants in two parties depending on the side of the bipartition they are in, and we add a priveleged partner in the other side to each participant, we prove that the accessing party has  to have a vertex from each pair of the matching : at least one person from each pair has to agree. Note that the minimal accessing sets are of size $2n+1$.
\TH
If $G$ is obtained from the complete bipartite graph $K_{n,n}$ by removing a matching $M=\{(u_i,v_i)\}_{i\in[1,n]}$, $G=K_{n,n}\setminus M$, $n \ge 1$. Then the accessing set for {\sf CC} secret sharing is defined as follows.
\AR{
Acc(G)&=&\{\{u_{i_1}\ldots u _{i_l}\} \cup \{ v_{j_1}\ldots v_{j_k}\}, k+l=n, \{ i_1 \ldots i_l\} \cup \{j_1 \ldots j_k\}= [1,n]\}
}
\HT
{\bf Proof.}
An accessing  set  is of the form $D \cup Odd(D)$ with $|D|$ odd. Thus $D$ contains an odd number of vertices in one side of the partition. Suppose that  $D \cup Odd(D)$ does not contain neither $u_i$ nor $v_i$ with $(u_i,v_i)$ an edge from the matching, then $D$ contains an even number of vertices in both sides of the partition as $u_i$ and $v_i$ are not in $Odd(D)$.
Furthermore, if a set contains exactly one vertex from each pair of the matching, it contains an odd number of vertices and thus it contains an odd number of vertices in one side of the bipartition the odd neighborhood of this part is exactly the set of vertices that are not linked to these vertices in the matching
so it is of the form $D \cup Odd(D)$ with $|D|$ odd.
\qed

In both these cases the conjugate graphs $G'$ have different access structures, so unfortunately they cannot be extended to {\sf CQ} and {\sf QQ} secret sharing. However this does not apply to the next example.

\begin{figure}[!h]
\begin{center}
\includegraphics[width=5cm]{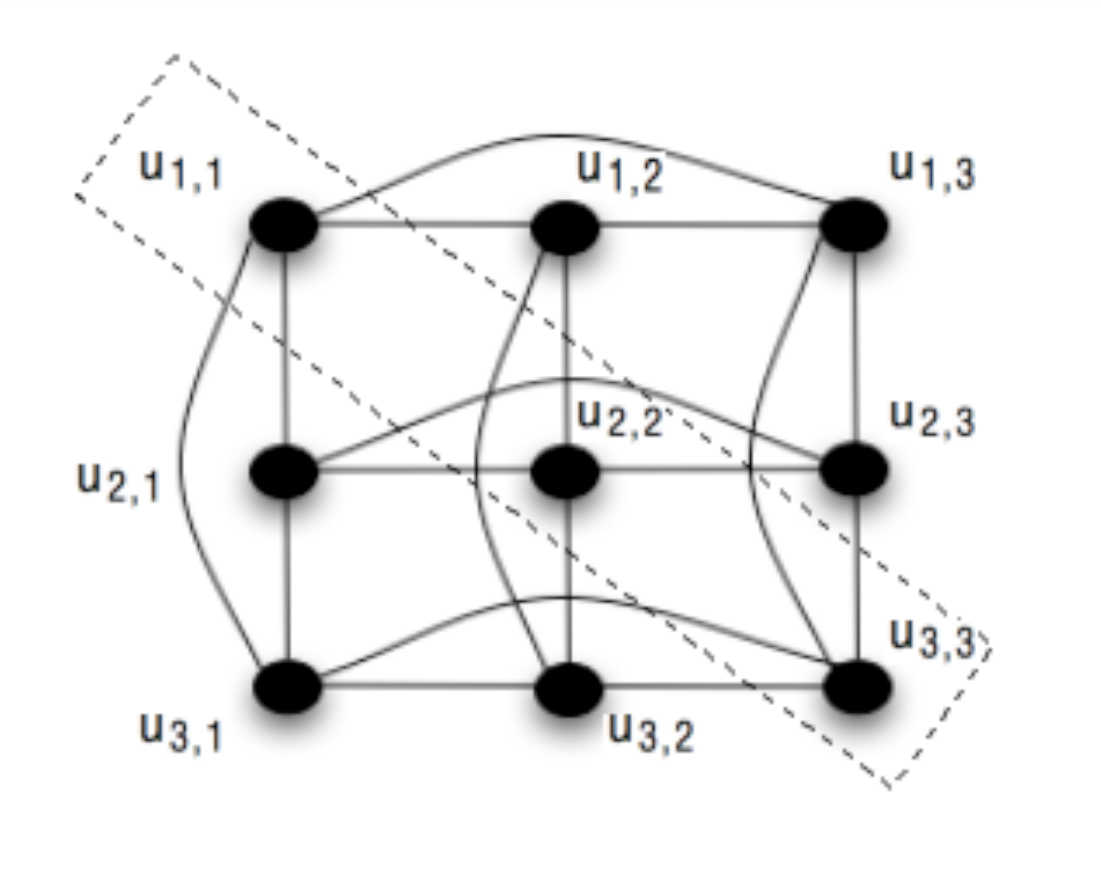}
\caption{Secret sharing using a torus.}
\label{tors}
\end{center}
\end{figure}

The final case is when $G$ is a $3\times 3$ torus  (cartesian product of two triangles $K_3\square K_3$, Figure \ref{tors}). We associate to each participant a line and a column and we prove that the accessing party has either one line and one column or one representative from each line and column.
Note that in this case, the property that minimal accessing sets are of same cardinality is relaxed.

\begin{lemme} \label{Lem: torus}
If $G=K_3 \square K_3$, then the accessing set for {\sf CC} secret sharing is the union of the diagonals and  neighborhoods  $Acc(G)=Dg\cup Ng$ where
\AR{
Dg&=& \{ \{u_{i_1,j_1},u_{i_2,j_2},u_{i_3,j_3} \},i_k\neq i_l \;\&\; j_k\neq j_l\} \\
Ng&=& \{ \{u_{i_1,j_1},u_{i_1,j_2},u_{i_1,j_3},u_{i_2,j_1},u_{i_3,j_1} \},i_k\neq i_l \;\&\; j_k\neq j_l \}
}
\end{lemme}
{\bf Proof.} 
We make use of the fact that the graph is vertex transitive  to simplify the case analysis.
 The sets $S=D \cup Odd(D)$ for $|D|=1$ are the neighborhoods $Ng$. If $|D|=3$ if $D$  has no edge then it is a diagonal $Dg$ and $Odd(D)=D$, if $D$ has one edge then $S$ contains a neighborhood, if $D$  has two or three edges then $S$ contains  a diagonal.If $|D|=5$ then if $D$ contains 3 vertices in a column and one vertex in the two others then it also contains a diagonal, if   $D$ contains 3 vertices in a column and 2 in an other column then $S=D \cup Odd(D)$ contains a neighborhood. If   $D$ contains 2 vertices in two first columns and 1 in  the last column then if the vertices in the first two columns are in the same lines then $S$ contains a neighborhood, otherwise $D$ contains a diagonal. If $|D|=7$ then D contains a diagonal.
\qed

The conjugate graph $G'$ is exactly the same form, but with vertices shuffled about - the diagonals of the original are now the rows and columns and vice versa. Explicitly the vertices $u_{i,j}'$ of $G'$ are given by 
$u_{1,1}'= u_{1,1}$, $u_{1,2}'= u_{2,2}$, $u_{1,3}'= u_{3,3}$, 
$u_{2,1}'= u_{3,2}$, $u_{2,2}'= u_{1,3}$, $u_{2,3}'= u_{2,1}$, 
$u_{3,1}'= u_{2,3}$, $u_{3,2}'= u_{3,1}$, $u_{3,3}'= u_{1,2}$. 

We can clearly see here that all the neighborhoods can access in both $G$ and $G'$, since the diagonals are not neighbours in one graph and so are clearly in the same neighborhood in the other graph. The same goes for neighborhoods. 

\begin{lemme}
If $G=K_3 \square K_3$, then for {\sf CQ} and {\sf QQ} secret sharing, all neighborhoods $Ng$ can access the secret.
\end{lemme}
{\bf Proof.} The proof follows from the construction of the conjugate graph and application of Lemma \ref{Lem: torus} to the conjugate graph. \qed

\bigskip

As well as these explicit examples, we can also make some more general statements about the implications of Theorems 1 and 2, which will help us see how far we can push these ideas. For example, it is known that the $(2,3)$-secret sharing protocol is impossible using these graph state schemes as noted in \cite{MS08}. In this direction, we can easily see the following two lemmas.

\begin{lemme}
For a {\sf CC} $(k,n)$-secret sharing scheme $(G,A)$, we must have $$|\enco| \geq n-k+1.$$
\end{lemme}
{\bf Proof.} By Theorem 2 set $S=V(G)/A$ cannot access the secret (indeed this is obvious since the reduced density matrix is independent of the secret). For the protocol to be $k$-accessible we thus demand that $k-1 \geq |S|=|V(G)/A|= n-A$ giving the lemma. \qed

\begin{lemme} \label{Lem: degree}
For a {\sf CC} $(k,n)$-secret sharing scheme $(G,A)$, we must have that the degree $deg(a)$ satisfies $$deg(a) \geq k-1$$ $\forall a \in A$.
\end{lemme}
{\bf Proof.} By Theorem 1, $a + N(a)$ can access the secret $\forall a \in A$. For the protocol to be $k$-private we thus demand that $deg(a) \geq k-1$ $\forall a \in A$.  \qed

Further application of these can get quite detailed, but we are able to make some general statements, for example the following lemma.

\begin{lemme}
For odd $n$, it is impossible to use graph state secret sharing \cite{MS08} to do a {\sf CC}  $(n-1,n)$-secret sharing protocol $(G,A)$ with $A=V(G)$. 
\end{lemme}
{\bf Proof.} To be an $(n-1,n)$ threshold scheme any set $S$, $|S|=n-1$ must be able to access the secret. From Theorem 2 we then have $N(a) \neq A \cap V / a$ $\forall a \in A$, which in turn implies that $deg(a) \neq |\cap V / a|$ $\forall a \in A$. When $A=V$ this reduces to $deg(a) \leq n-2$ $\forall a \in V$. Along with Lemma \ref{Lem: degree} we have that $deg(a) = n-2$ $\forall a \in V$, which is impossible to satisfy for $n$ is odd.

\section{Flow of Secret Sharing}

Generally speaking, one can view a {\sf QQ} quantum secret sharing as a one-way pattern where the dealer possess the input qubit, the output is one of the qubit in the access set, and all the qubits of unauthorised parties (who cannot access the secret) can be viewed as \emph{pointless measurements}, \ie~measurements that have no effect on the deterministic computation from input to output.  Recall that we can characterise a deterministic computation from input to output in a given graph using flow or generalised flow \cite{Flow06, g-flow}. In this section we present a set of simple rules that graphically characterises a pointless measurement with respect to a given flow construction. This will be the basis of our scheme for extending a given one-way pattern into a secret sharing protocol.



%


For a given pattern ${\mathcal P}=(V,I,O,A)$ and an auxiliary qubit $u\in V\setminus (I\cup O)$, we define ${\mathcal P}|_u$ to be a new pattern where the measurement of $u$ is removed, \ie~$u$ is no longer measured and it becomes an output qubit, and hence the signal $s_u$ is set to $0$.


\DE
For a given pattern ${\mathcal P} = (V,I,O,A'M_u^\alpha A)$, let $${\mathcal P}|_u :=(V,I,O\cup\{u\}, (A'[s_u\leftarrow 0])A)$$
\ED

Let $F$ be a completely positive trace preserving map acting on a space including qubit $u$, then the tracing out operator $tr_u(F)$, is defined as
\AR{
\rho\mapsto tr_u(F(\rho)) := \rho\mapsto \sum_{x\in \{0,1\}} \bra x_uF(\rho)\ket x_u
}



\DE
We define the measurement of qubit $u$ in pattern $\mathcal P$ to be \emph{pointless} if  $\den{\mathcal P}=tr_u(\den{\mathcal P|_u})$, where $tr_u(.)$ is the trace out of the qubit $u$.
\ED

The trace command, $tr_u$, can be easily encoded in the measurement calculus, i.e. for any pattern $\mathcal P = (V,I,O,A)$, there exists
$\mathcal P'=(V,I,O\setminus \{u\}, A')$ s.t.
\AR{
tr_u(\den {\mathcal P}) =\den {\mathcal P'}
}

\LE \label{l-traceout}
For any pattern $\mathcal P = (V,I,O,A)$ and any $u\in I^c$, if
\AR{
\mathcal P\Rightarrow^* (V,I,O,A_CA_MA_E)
}
then
\AR{
tr_u(\den {\mathcal P}) =\den {\mathcal P'}
} where $\mathcal P'=(V,I,O\setminus \{u\},A_CA_M(\prod_{v\in N(u)} Z_v^{s_u} )M_u^ZA_E)$.
\EL


In the following theorem, we prove that the accessibility in a secret sharing protocol with graph states is related to the existence of a pattern $\mathcal P$  where all the qubits in $V\setminus S$ are pointless.

\TH For a given {\sf CC} secret sharing protocol $G$, with secret encoded in $A$ and a given $k \ge 0$, if for any $S\subseteq V(G)$ s.t. $|S|=k$ there exists a pattern $\mathcal P$ s.t.
\begin{itemize}

\item The underlying graph of $\mathcal G(P)$ is $G$ augmented with an input vertex $a$ connected to all the vertices of $A$, and where the output vertex is any vertex of $S$. 
\item $\den {\mathcal P}$ is unitary ;
\item ${V\setminus S}$ is pointless in $\mathcal P$
\end{itemize}
then the protocol is a $(|V(G)|,k)$- {\sf QQ} quantum secret sharing protocol.
\HT

{\bf Proof.}
Any set of $k$ partners $S$ can apply the pattern given by the theorem to compute $\den {\mathcal P}$ independently of the action of the others. Thus any $k$ partners can access the information, just by applying $\den {\mathcal P}^\dagger$ on the output qubit. Any fewer than $k$ partners cannot since it is already a {\sf CC} secret sharing protocol, and if they cannot access the classical information, their reduced states are independent of the quantum information also, hence it is inaccessible.
\qed








Although it seems a hard task to characterise which qubits might become pointless the following result gives us a way to handle some special cases. More precisely we present a necessary and sufficient condition for a qubit being pointless with respect to a given flow construction and assuming that all vertices are measured in the $(X,Y)$ plane. The intuition behind the theorem is to find a set of conditions that makes the $Z$ and $X$ errors induced by tracing out the pointless measurement, ineffective. In order to analyse the $Z$ error propagation we will use the notion of \emph{influencing walks} for geometries \cite{BK06}.

\begin{defi} Let $(f,\preceq)$ be the flow of a geometry
$(G,I,O)$. An \emph{influencing walk} at node $v$ is an~$I-v$ walk in $G$ that starts with a flow
edge, has no two consecutive non-flow edges and traverses flow edges in the forward direction.
\end{defi}

And to analyse the effect of an $X$ error we will use the phase map decomposition for unitaries \cite{BDK06} which leads to a simple characterisation of patterns implementing the same unitary. Assuming that $|V|=m$ and $|I|=|O|=n$, we define the index set $A_{pq}$ for $p,q \in \{0,1\}^{n}$ as follows
\AR{
A_{pq}=(p+k(2^n - 1), p+(k+1)(2^n - 1), \cdots, p+k'(2^n - 1))
}
where $k$ is the smallest integer such that
\AR{
p+k(2^n - 1) \ge (q-1) 2 ^{m-n} + 1
}
and $k'$ is the largest integer such that
\AR{
p+k'(2^n - 1) \le q 2 ^{m-n}
}

\begin{figure}[!h]
\begin{center}
\includegraphics[width=7cm]{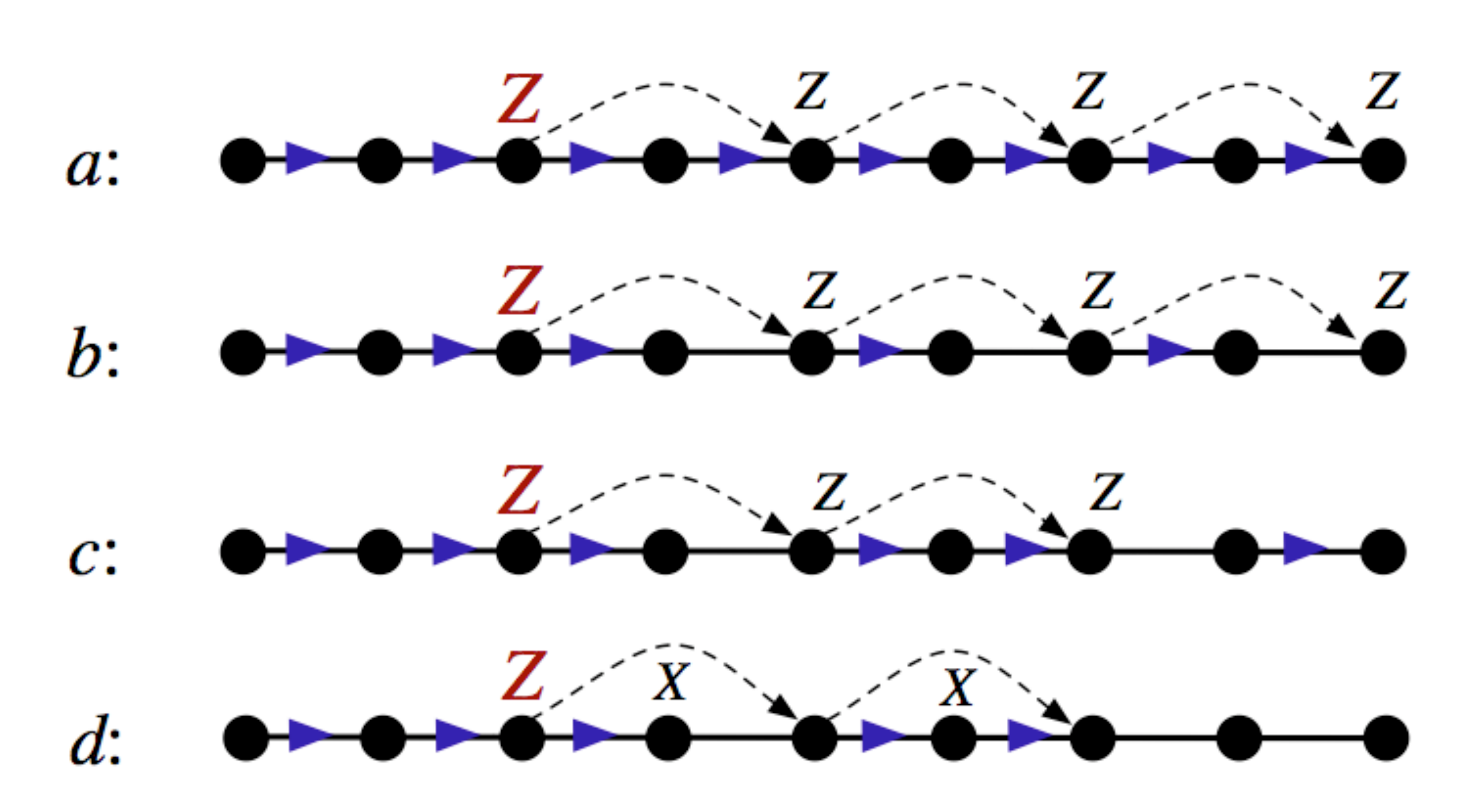}
\caption{Propagation of $Z$ errors}
\label{f-Zprop}
\end{center}
\end{figure}

\TH
Let $\mathcal P$ be a pattern on graph $G(V,I,O)$ with angles of measurement $\{\al_i\}_{i\in O^C}$ and let $p$ be a qubit in $V$ such that the graph $G\setminus p$ has flow. The measurement at qubit $p$ is pointless if and only if the following conditions are satisfied
\begin{itemize}
\item[(a)] $\exists S \subset V\setminus O$, $S\neq \emptyset$, such that $\odd(S)\subseteq N_G(u)$.
\item[(b)] For all $p,q \in \{0,1\}^{|I|}$ we have
\AR{
\sum_{x\in A_{pq}} \emi{\sum_{O^c}\al_jx_j} = \sum_{x\in A_{pq}}\emi{\sum_{O^c}\al'_jx_j}
}
where
\AR{
\al'_j = \left\{
\begin{array}{l}
-\al_j \;\;\;\;j \in S\\
\al_j \;\;\;\; \text{otherwise}
\end{array}
\right.
}
\end{itemize}
\HT
{\bf Proof.} First, we discuss the forward direction, \ie~we assume $p\in V$ is a pointless measurement and we show how the above conditions are satisfied. We know from Lemma \ref{l-traceout} that tracing out $p$ introduces $Z$ errors on all qubits in $N(p)$ however these errors do not change the underlying computation of the pattern. Recall that a $Z$ error on a qubit will flip the result of measurement on that qubit ($F_i$)
\AR{
\M \al i \cz i {} = \M {\al+\pi} i = F^1_i \M \al i
}
Therefore the $Z$ errors resulting from the trace out operator, will propagate through pattern and will have an effect on the final output. To analyse their effect we note that $Z$ errors can only propagate through  an influencing walk over $G \setminus p$ which has a unique flow. This is due to the fact that any two dependent commands lay on a same influencing walk \cite{BK06}. Furthermore the structure of flow dependencies along a walk are in such a way that a $Z$ error will propagate over even distanced qubits and hence introduces new $Z$ errors at these even distanced qubits, Figure \ref{f-Zprop} part $a$ and $b$. However a $Z$ error will not propagate on a walk with a non-flow edge starting at even distance of the $Z$ error, Figure \ref{f-Zprop} part $c$. Finally qubit at odd distance from a $Z$ error that has been propagated will receive an $X$ error, Figure \ref{f-Zprop} part $d$.
%

On the other hand since $p$ is pointless, on any output qubits the above induced errors must cancel out. Therefore a $Z$ error at qubit $x\in N(p)$ (induced due to tracing out $p$) on a walk starting at $x$ ending at output $o$ (that introduce either $X$ or $Z$ error on $o$) has to be cancel out by
\begin{itemize}
\item [(i)] either hitting another $y_x\in N(p)$ at an even distance on the same walk
\item [(ii)] or there exists another unique walk also ending at $o$ with $y_x\in N(p)$ such that $x$ and $y_x$ are even distanced over the walk $x-c-y_x$, where $c$ is the first common qubit on both walks.
\end{itemize}

We can now construct a set $S$ such that $\odd(S)\subseteq N_G(u)$. Let for any $x\in N(p)$, $S_x$ to be the set of qubits between $x$ and $y_x$ over the common walk for the case (i) and over the walk $x-c-y_x$ for the case (ii), and at odd distances from $x$ and define
\AR{
S=\cup_{x\in N(p)}S_x}
In order to show
\AR{
Odd(S) \subset N(p)
}
we prove for any $l\not \in N(p)$ and any of its neighbour $s\in S$ there exists another of its neighbour $s'\in S$, which implies $l\not \in Odd(S)$. Due to flow construction, there must exists an $x\in N(p)$ such that $s \in S_x$ and the incoming edge into $s$ in $S_x$ is a flow edge. Hence $x$ is at even distance from $l$ and the $Z$ error at $x$ can propagate along $x-l$. Therefore there exists a $y_x$ at even distance from $l$ along the walk starting with $x-l$, which implies the neighbour of $l$ along $x-y_x$ belongs to $S$.

So far we have ignored the $X$ error that has been produced due to the propagation of the $Z$ error from $x$ into $y_x$, that is the $X$ errors over the set $S$. However unlike a $Z$ error, an $X$ error cannot be propagated and will have a direct effect on the measurement of underlying qubit
\AR{
\M \al i \cx i {} = \M {-\al} i
}
Let $\{\al'_i\}$ to be as defined in the theorem, since $p$ is a pointless measurement, the two patterns over $G\setminus p$ with angles $\{\al_i\}$ and $\{\al'_i\}$ should implement the same unitary which from phase map decomposition leads to the statement of condition $b$ in the theorem.

The proof in other direction is straightforward, since due to the above discussion the existence of a set $S$ will permit one to discharge the effect of all induced $Z$ errors obtained from tracing out qubit $p$ which concludes $p$ has a pointless measurement.
\qed

One simple scenario for condition $(b)$ is when all qubits in $S$ are measured with Pauli $X$. One can easily extend the above theorem to characterise a set of vertices that are simultaneously pointless with respect to a given pattern with flow. However a further adjustment is required to extend the same approach to the domain of generalised flow \cite{g-flow}.




The first step in using the above theorem is to start with a fixed graph with input and output with flow, then by adding appropriate measurements and edges we extend the graph to one with maximum number of pointless measurements. However via this method one might not obtain a symmetric protocol, that is a pattern where we can choose which qubits to become pointless after choosing the output qubits. We leave as an open question how one could use the above result for building a true secret sharing protocol.

\section{Conclusions}
In this work we have considered graph state secret sharing protocols of \cite{MS08}, their complete graphical description and formal connection to other one-way quantum information protocols.

We have given graphical rules on which sets of players can access the secret, and which cannot. These rules were then used to generate and prove the validity of new protocols, as well as make some general statements and no-goes for graph state protocols. Their full extent and use for further protocols and no goes remains open.

We have also given necessary and sufficient conditions for the addition of vertices corresponding to pointless measurements (i.e. measurements which do not effect the outcome of a computation) in a one way pattern with flow. In the context of secret sharing this offers a path to generate new quantum secret sharing protocols directly. This result also has interest in the context of fault tolerance where it can be viewed as offering a path to allow for faulty qubits.

Both of these sets of contributions put secret sharing closer to the general framework of measurement based quantum information processing and, as such, open up the possibility to incorporate it into larger network tasks involving several different elements such as computation, error correction and secret sharing. In the other direction, the techniques and results developed may prove useful beyond secret sharing in particular for the development fault tolerant MBQC.

\section*{Acknowledgments}
DM is acknowledges financial support from the French ANR Defis program under contract
ANR-08-EMER-003 (COCQ project).


\bibliographystyle{eptcs}

\end{document}